\documentclass[aps,pre,preprint]{revtex4}

\usepackage{graphicx}
\usepackage{dcolumn}
\usepackage{bm}
\usepackage{amsmath}
\usepackage{amssymb}
\usepackage{color}
\usepackage{tikz}

\usepackage{hyperref}

\definecolor{darkred}{rgb}{0.5,0,0}

\begin{document}

\title{Effect of the density of pillar-patterned substrates on the contact mechanics: Transition from top to mixed contact with a detailed pressure-field description}

\author{René Ledesma-Alonso}
 \email{rledesmaalonso@gmail.com}
 \affiliation{Departamento de Ingenier\'{i}a Industrial y Mec\'{a}nica, Escuela de Ingenier\'{i}a, Universidad de las Am\'{e}ricas Puebla, San Andr\'{e}s Cholula, Puebla, C.P. 72810, M\'{e}xico.}
\author{Elie Raphael}
\affiliation{ Laboratoire de Physico-Chimie Th\'{e}orique, UMR CNRS/ESPCI Gulliver 7083, 75005 Paris, France.}%
\author{Frédéric Restagno}
 \affiliation{Universit\'{e} Paris-Saclay, CNRS, Laboratoire de Physique des Solides, 91405 Orsay cedex, France.}%
\author{Christophe Poulard}
 \affiliation{Universit\'{e} Paris-Saclay, CNRS, Laboratoire de Physique des Solides, 91405 Orsay cedex, France.}%

\date{\today}

\begin{abstract}
Different contact regimes between a spherical lens and a periodically patterned substrate are observed, when they are pressed against each other.
Top contact occurs when only the highest substrate sections touch the lens, whereas mixed contact implies that both the highest and the lowest substrate sections come into contact with the lens.
In this paper, we study how the pattern density of the substrate, along with its physical properties and those of the lens, determine the transition from top contact to mixed contact.
Experiments and numerical simulations had been performed, as complementary approaches to obtain data, and a theoretical analysis has been developed to gain insight on the effect of the physical parameters on the contact transition.
As a result, a phase diagram is presented, in terms of the load and the contact radius, that combines the observations of the three approaches (experimental, numerical, and theoretical), unveiling the boundaries of three contact regimes: (1) deterministic-driven contact, (2) top contact, and (3) mixed contact. \\[2em]
Cite this article: \\[1em]
R. Ledesma-Alonso, E. Raphael, F. Restagno, and C. Poulard, Phys. Rev. E 104(5), 055007
(2021). \\
DOI: \href{https://link.aps.org/doi/10.1103/PhysRevE.104.055007}{10.1103/PhysRevE.104.055007}
\end{abstract}

\pacs{Materials physics - Mechanical \& acoustical properties - Mechanical deformation - Elastic deformation}
\maketitle

\section{Introduction}

The problem of the contact of elastic bodies with nominally flat surfaces is a classical problem of continuum mechanics, which has been first proposed by Heinrich Hertz in a paper submitted to the Berlin Physical Society in 1881~\cite{Hertz1882,Johnson1982}. 
However, practical materials surfaces are not flat and present roughnesses on a wide range of length scales.
A consequence of the roughness is that the real contact area between two surfaces is usually much smaller than the apparent contact area.
This imperfect contact has profound implications for transmission of charge, heat, and forces through the interface.
In particular, the reduction of the contact area allowed to explain one of the nonintuitive Coulomb-Amontons laws of friction: The friction coefficient between two solids is independent of the normal load~\cite{Bowden,Dowson}.
For metallic surfaces Bowden and Tabor proposed that the contact between two metals could be determined by assuming a fully plastic deformation of the junction asperities in contact~\cite{Bowden}.
Later, Archard~\cite{Archard1953}, Greenwood and Williamson~\cite{Greenwood1966}, and Bush {\em et. al.}~\cite{Bush1975} pioneered the development of models for contact between complex elastic surfaces.

The case of the contact of surfaces presenting a self-affine fractal character has been first developed by Tossati and Persson~\cite{Persson2001}, providing a good theoretical model to describe real surfaces, since it is well supported by several experimental studies~\cite{Krim1993,Palasantzas1993,Bouchaud1997,Jacobs2017}.
From this pioneering work several theoretical developments have been done to describe different contact properties~\cite{Persson2002,Bhushan2004,Dalvi2019}, or to combine elastic and plastic deformation~\cite{Pei2005,Pastewka2013}.
A modern application of this complex contact mechanics description is haptic systems, which allow users to ``feel'' virtual objects in a simulated environment~\cite{Adams1999,Hayward2004,Salisbury2004}.

Nevertheless, more simple systems than fractal surfaces have also been employed to understand the effect of roughness on contact problems, for instance surfaces with periodic patterns~\cite{Westergaard1939,Johnson1984,Crosby2005,Block2008,Hui2011,Poulard2011}.
A potential application of these patterned surfaces comes from the increasing interest to use them as biomimetic surfaces~\cite{Ghatak2004,Hui2004,Benz2006,Kim2007,Zeng2009,Varenberg2009,Bartlett2012,Das2013,Nguyen2013,Brodoceanu2016}.
From this point of view, one of the important question is the contact formation, which fixes the relationship between the measured adhesion and the preload before detachment~\cite{Benz2006}.
Additionally, the problem of the dynamical impact of a solid sphere onto a textured elastic surface has been studied~\cite{Maruoka2019}, which is an example of other related physical phenomena that have not yet been explored in depth.

Remaining on the track of simple systems, the contact between a spherical lens on a well-controlled patterned surface made on elastomeric Polydimethylsiloxane PDMS with periodic circular bumps or cylindrical pillars is an interesting contact model.
The geometry and elastic properties of the textured surface (array of bumps or pillars over a flat substrate), and its effect on the contact formation, have been studied experimentally~\cite{Crosby2005,Verneuil2007,Hisler2013,Poulard2013,Poulard2015}.
Interestingly, when the two surfaces are pushed together, a transition from top contact (the lens touches only the top of the bumps or pillars) to full or mixed contact (the lens gets in contact not only with the bumps or pillars, but also with the underlying substrate in between the patterns) has been observed.
A clear experimental dependence of the transition conditions on the bumps or pillars density, has been found for this discrete contact phenomenon~\cite{Verneuil2007,Poulard2013,Poulard2015}.

To describe their experiments, Verneuil~{\em et. al.}~\cite{Verneuil2007} developed an initial model based on a continuous Hertzian distribution for the stress inside the contact region, allowing them to determine a critical force $F_c\propto \phi^3$, where $\phi$ is the surface fraction covered by bumps or pillars.
Poulard~{\em et. al.}~\cite{Poulard2013} complemented this model by including the coupling between pillars at small separation distances, which leads to a better representation of the data for large values of $\phi$.
With a theoretical approach, Ledesma-Alonso~{\em et. al.}~\cite{Ledesma2016} calculated the exact stress distribution inside the discontinuous region of contact, for any value of $\phi$.
Their results showed that for $\phi \rightarrow 1$, a Hertzian-like stress distribution could be used to describe the phenomenon due to small distance between pillars, whereas for $\phi \rightarrow 0$, with the pillars being far apart, the contribution of a Boussinesq-Cerruti-like stress distribution at each pillar has to be considered.

In this paper, we study the transition from top contact to mixed contact that occurs when a spherical lens and a patterned surface (pillar lattice on a flat substrate), both being elastic, are compressed against each other.
We analyze the parameters that provoke the transition phenomenon, focusing on the force, contact radius and indentation.
The effect of the surface fraction (pillar density) on these parameters is studied by performing experiments and numerical simulations.
The results of these two approaches are compared, along a wide range of the surface fraction values.
Additionally, a theoretical analysis is presented, from which small and large contact radius approximations can be deduced.
These limit cases provide a good description of the general trends that the experimental and numerical simulation results follow. 
The theoretical expressions of the force, contact radius and indentation, for the small and large contact radius approximations, are considerably simple, despite the geometrical complexity of the mechanical contact between the lens and the patterned surface.
Finally, a phase diagram of the contact regimes is depicted in terms of the force and contact radius, valid for different orders of magnitude of the surface fraction.

\section{System description}

\begin{figure}
\centering
\includegraphics[width=0.58\textwidth]{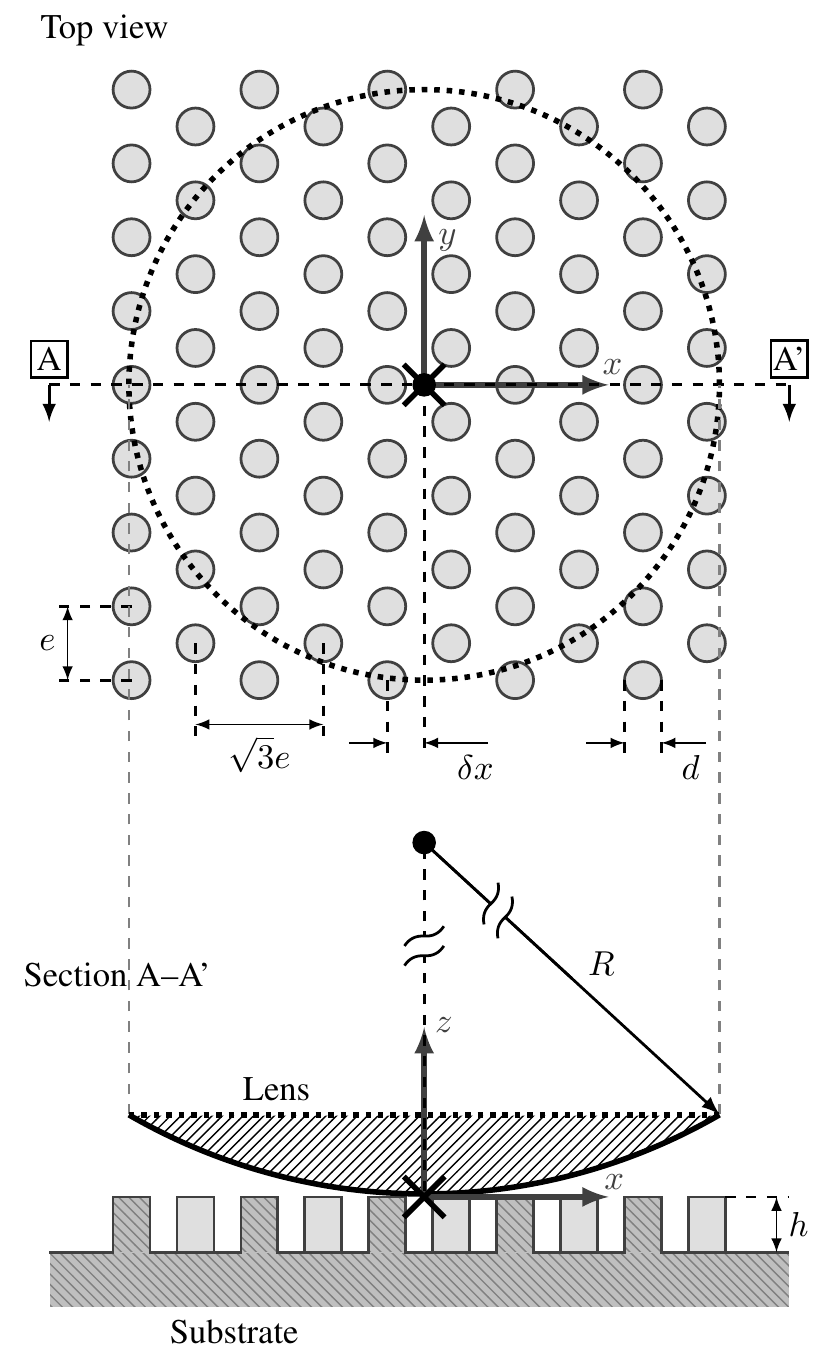}
\caption{
Schema with the variables that represent the characteristic dimensions of the physical system.
The black cross {(\boldmath$\times$)} indicates the target position and the origin of coordinates.
The dotted line ({\protect\tikz[baseline=-0.6ex]{\protect\draw[ultra thick,black!100!white,dotted](0,0) -- (0.6,0);}}) indicates the horizontal boundary of the lens for the top view, and the upper boundary for the Section A--A'.
}
\label{Fig:Schema}
\end{figure}

Consider a lower spherical cap or lens of radius $R$ and a periodically rough substrate, composed by an hexagonal lattice of cylindrical pillars, each one of diameter $d$ and height $h$, placed atop a flat semi-infinite body, and separated by a pitch $e$ between the center of two pillars.
Hereafter, the part of the substrate that is covered by the pillars will be called \emph{top substrate}, whereas the part of the substrate that is not covered by the pillars will be called \emph{bottom substrate}.
The origin of coordinates is placed below the lowest point of the lens, at a target position atop the textured substrate, as it is depicted in Fig.~\ref{Fig:Schema}, with the $x$ and $y$-directions contained in the horizontal plane and the $z$-direction pointing upwards.
For an hexagonal lattice of pillars, we can define the fraction of the surface occupied by the pillars as:
\begin{equation}
\Phi=\dfrac{\pi}{2\sqrt{3}}\left(\dfrac{d}{e}\right)^2 \ .
\label{eq:phi}
\end{equation}
As shown in Fig.~\ref{Fig:Schema}, the initial position of the lens $f_l^{\circ}$ reads:
\begin{equation}
f_l^{\circ}\left(x,y\right)=h+R-\sqrt{R^2-\left(x^2+y^2\right)} \ ,
\label{eq:lens0}
\end{equation}
whereas the initial shape of the textured substrate $f_s^{\circ}$ is described by:
\begin{equation}
f_s^{\circ}\left(r\right)=h\sum_{n=1}^N\left[1-H\left(r_n-\dfrac{d}{2}\right)\right] \ .
\label{eq:subs0}
\end{equation}
In the last expression, $H$ is the Heaviside step function
and $r_n$ is a relative distance, given by:
\begin{equation}
r_n\left(x,y\right)=\sqrt{\left(x_n-x\right)^2+\left(y_n-y\right)^2} \ ,
\label{eq:radd}
\end{equation}
measured between the coordinates $(x,y)$ and the position of the center of the $n$th pillar $\left(x_n,y_n\right)$.

As usual, the elastic properties of the materials that constitute the system must be considered, since the mechanic interaction between the involved bodies depend on their magnitudes.
For each body $i$, being $i=l$ for the lens and $i=s$ for the textured substrate, its Young's elastic modulus $E_i$ and its Poisson's ratio $\nu_i$ are evoked.

\begin{figure}
\centering
\includegraphics[width=0.32\textwidth]{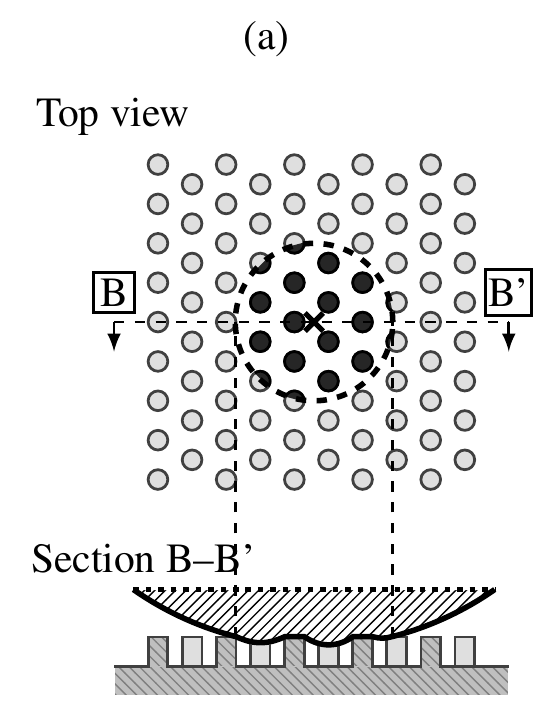}
\includegraphics[width=0.32\textwidth]{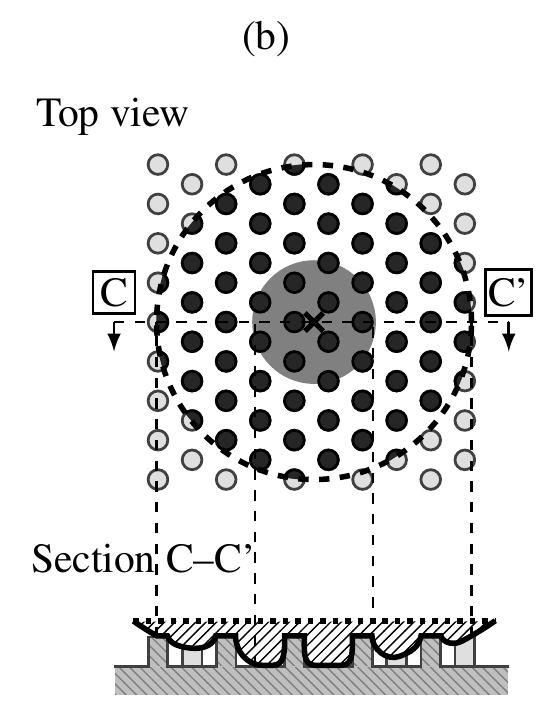}
\caption{
Top and section view of the system with deformed geometries, due to the application of a force with different magnitudes, leading to: (a) top contact and (b) mixed contact.
System layout is the same as in Fig.~\ref{Fig:Schema}.
At the top view, dark gray indicates the contact between the lens and the top of the pillars, for both top and mixed contact, as observed in the section B--B' and Section C--C' views.
For mixed contact, the circular gray region around the target position {(\boldmath$\times$)} indicates the contact between the lens and the underlying substrate in between the pillars, as observed in the Section C--C' view.
}
\label{Fig:Transition}
\end{figure}

When the two bodies are compressed against each other with a force $F>0$, coupled with an indentation $\zeta>0$, the contact region extends from the origin of coordinates to a contact radius $a$, and the surfaces of both bodies suffer a significant deformation inside and outside the contact region.
Depending on the magnitude of $F$, two different contact regimes may occur: (1) top contact and (2) mixed contact, for magnitudes smaller or larger than a threshold or critical load $F_c$, respectively.
Top contact implies that the lens touches only the top of the pillars, whereas mixed contact implies that the lens gets in contact not only with the pillars, but also with the underlying substrate in between the pillars.
The corresponding values of the contact radius and the indentation, associated with the critical load $F_c$, are also defined as the critical contact radius $a_c$ and the critical indentation $\zeta_c$.
Both regimes are depicted in Fig.~\ref{Fig:Transition}.
Top contact consists in the contact between the top substrate and the lens, within the contact region.
Mixed contact consists in a combination of top contact, within a large region, and an additional contact between the lens and the bottom substrate, within a small region close to the origin of coordinates.

In this work, we are interested in the transition from top to mixed contact, particularly on the threshold conditions: force, contact radius, and indentation.
To achieve this objective, an experimental approach, a numerical method and a theoretical formulation had been developed.
In all cases, a top contact situation is initially considered, but the indentation is gradually increased until the conditions for which the very first mixed contact occur, \emph{i.e.}, a zero gap between the bottom substrate and the lens at some localized positions $(x_c,y_c)$ takes place, just before the mixed contact interplays with the two bodies deformation.

\subsection{Starting from a top contact}

For a top contact, the gap between the two surfaces is obtained with the following expression:
\begin{equation}
\Delta\left(x,y\right)=f_l\left(x,y\right)-f_s\left(x,y\right) \ ,
\label{eq:gap}
\end{equation} 
where the position of the lower surface of the lens $f_l$ and that of the upper surface of the textured substrate $f_s$ are given by: 
\begin{subequations}
\begin{align}
f_l\left(x,y\right) &=f_l^{\circ}\left(x,y\right)+\sum_{n=1}^N w_{l,n}\left(x,y\right)-\zeta \ ,\\
f_s\left(x,y\right)&=f_s^{\circ}\left(x,y\right)-\sum_{n=1}^N w_{s,n}\left(x,y\right) \ .
\end{align}
\label{eq:posls}
\end{subequations}
where $w_{l,n}$ and $w_{s,n}$ represent the displacement fields at the surface of the lens and the textured substrate, respectively, induced  by the contact between the lens and the top of the $n$th pillar, with a total of $N$ pillars in contact.
The indentation is thus represented by $\zeta$.

Consequently, the gap is given by:
\begin{equation}
\Delta\left(x,y\right)=f_l^{\circ}\left(x,y\right)-f_s^{\circ}\left(x,y\right)+w\left(x,y\right)-\zeta \ ,
\label{eq:ind}
\end{equation} 
where $w$ is the total displacement field, which reads:
\begin{align}
w\left(x,y\right)=\sum_{n=1}^N \left[w_{l,n}\left(x,y\right)+w_{s,n}\left(x,y\right)\right] \ .
\label{eq:totdisp}
\end{align}
It is important to remark that within the contact region, which corresponds to the top surface (either complete or partial) of the pillars in contact with the lens, the gap is $\Delta(x,y)=0$.

The compression force is computed as follows:
\begin{equation}
F=\int_0^{2\pi}\int_0^{a} \sigma\left(r,\theta\right)\,r\, drd\theta \ ,
\label{eq:force}
\end{equation}
where $r=\sqrt{x^2+y^2}$ and $\theta=\tan^{-1}(y/x)$ are respectively the radial and angular coordinates, and $\sigma$ is the stress field over the contact region, which is related to the total displacement field $w$, and demarcated by the contact radius $a$.
A relationship between $w$ and $\sigma$, which involves the elastic properties $E_i$ and $\nu_i$ of the lens and the substrate, is required in order to solve the problem, leading finally to the identification of $F$ and $a$ for an imposed indentation $\zeta$.

\subsection{Transition from top to mixed contact}

For relatively small indentations only top contact is observed, but when the indentation is increased, the transition from top to mixed contact occurs when the critical indentation $\zeta_c$ is reached.
Above the critical value $\zeta_c$, only mixed contact takes place.
In order to reach the precise value of $\zeta_c$, a specific compression force should be applied, known as the critical load $F_c$.
Besides the critical indentation and load, there is also another value that arises during the transition, which is the critical contact radius $a_c$.
Within the entire contact area defined by the critical radius $a_c$, top contact takes place, whereas at some specific positions defined as $(x_c,y_c)$, the mixed contact is triggered.
At the positions $\left(x_c,y_c\right)$, the gap between the bottom substrate and the lens becomes $\Delta\left(x_c,y_c\right)=0$.

Therefore, the transition from top to mixed contact is characterized by the simultaneous occurrence of the critical values $F_c$, $a_c$, and $\zeta_c$, and, as it happens for the Hertzian contact, the three quantities are coupled and described by simple relationships, as it will be detailed in the following section.

\section{Resolution approaches}

\begin{table}
\caption{Geometric and elastic parameters of the lens.}
\small
\begin{tabular}{|c|c|c|}
\hline
Radius & Poisson's ratio & Young's modulus \\
$R$, [$\mu$m] & $\nu_l$, [1] & $E_l$, [MPa] \\ \hline
$2200$ & $0.5$ & $2.0$ \\ \hline
\end{tabular}
\label{Tab:Lens}
\end{table}

\begin{table}
\caption{Geometric and elastic parameters of the textured substrate.}
\small
\begin{tabular}{|c|c|c|c|c|}
\hline
Height & Diameter & Poisson's & Young's & Relative \\
& & ratio & modulus & condition \\
$h$, [$\mu$m] & $d$, [$\mu$m] & $\nu_s$, [1] & $E_s$, [MPa] & \\ \hline
& & & $1000$ & rigid \\ \cline{4-5}
$2.2$ & $6.0$ & $0.5$ & $2.0$ & intermediate \\ \cline{4-5}
& & & $0.002$ & soft \\ \hline
\end{tabular}
\label{Tab:Subs}
\end{table}

\begin{figure}[h]
\centering\includegraphics[width=0.6\textwidth]{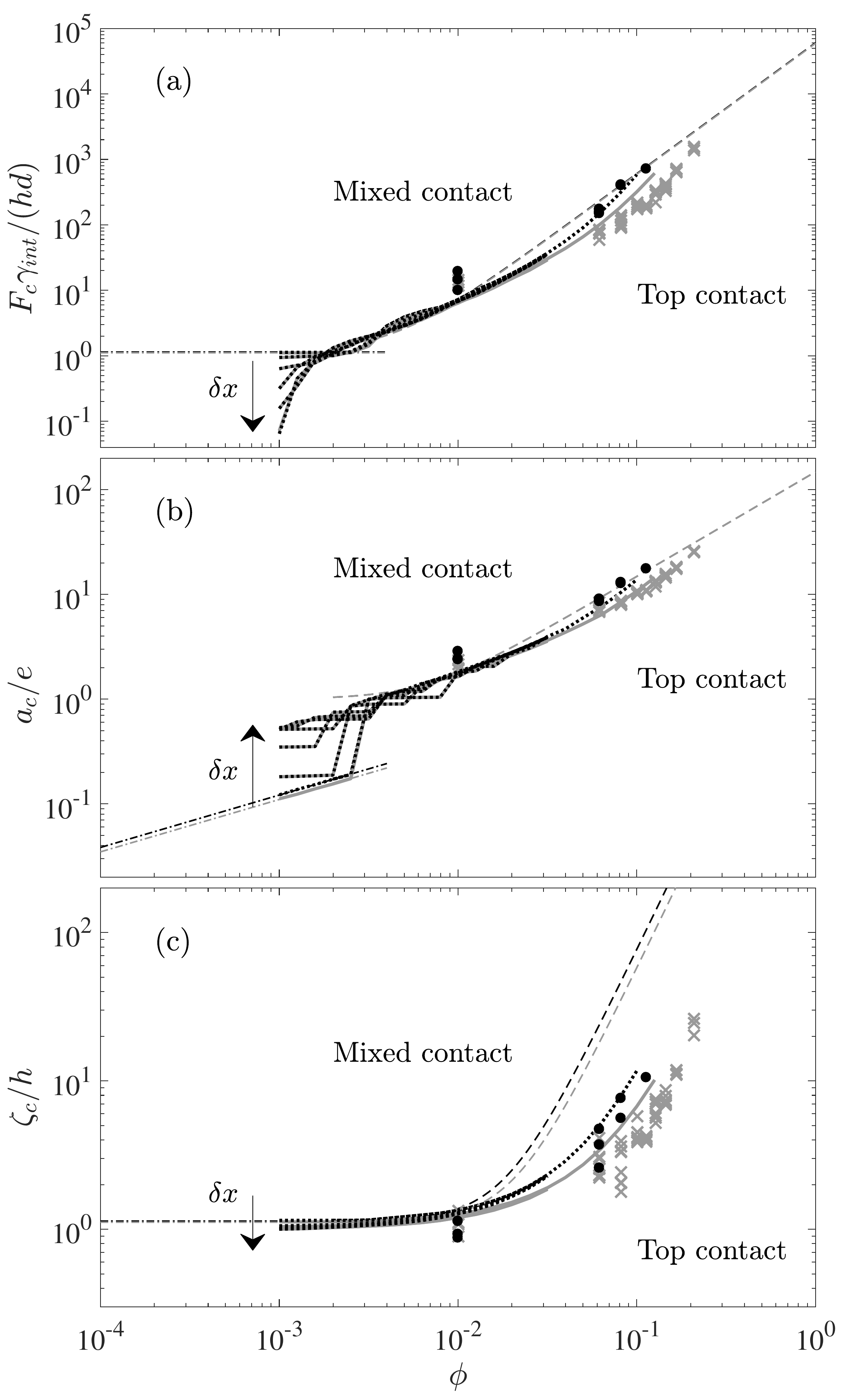}
\caption{
(a) Critical force $F_c$, (b) contact radius $a_c$, and (c) the required indentation $\zeta_c$ to transit from top to mixed contact, as functions of the surface fraction $\phi$.
Experimental ($\bm{\times},\bm{\bullet}$) and numerical simulation ({\protect\tikz[baseline=-0.6ex]{\protect\draw[black!100!white,line width=1.2pt](0,0) -- (0.6,0);}},{\protect\tikz[baseline=-0.6ex]{\protect\draw[black!100!white,dotted,line width=1.2pt](0,0) -- (0.6,0);}}) results are depicted, for intermediate (gray,$\bm{\times}$,{\protect\tikz[baseline=-0.6ex]{\protect\draw[black!100!white,line width=1.2pt](0,0) -- (0.6,0);}}) and rigid (black,$\bm{\bullet}$,{\protect\tikz[baseline=-0.6ex]{\protect\draw[black!100!white,dotted,line width=1.2pt](0,0) -- (0.6,0);}}) conditions of the textured substrate, according to the parameters given in Tables~\ref{Tab:Lens} and \ref{Tab:Subs}.
Simulations results are represented, corresponding to different target positions $\delta x$ atop the textured substrate.
The arrows indicate increasing values of $\delta x$ from 0 to $(\sqrt{3}/2)e$.
Analytic solutions are also depicted for intermediate (gray) and rigid (black) conditions: Dash-dotted curves ({\protect\tikz[baseline=-0.6ex]{\protect\draw[black!100!white,line width=1.2pt,dash dot](0,0) -- (0.6,0);}}) are Eqs.~\eqref{eq:small}, for a single pillar ($a_c<e$, in Sec.~\ref{sec:small}); dashed curves ({\protect\tikz[baseline=-0.6ex]{\protect\draw[black!100!white,line width=1.2pt,dashed](0,0) -- (0.6,0);}}) are Eqs.~\eqref{eq:large}, for $N>1$ pillars ($a_c>e$, in Sec.~\ref{sec:large}).
}
\label{Fig:FaZPhi}
\end{figure}

The values of the geometric and physical parameters of the system, conformed by the spherical lens and the textured substrate, that had been considered in this study, are presented in Tables~\ref{Tab:Lens} and \ref{Tab:Subs}.
As it has been  already mentioned, three approaches (experimental, numerical and theoretical) had been developed to study the transition from top to mixed contact, and to find the relations between the threshold conditions (force, contact radius, and indentation).

\subsection{Experiments}

The experimental set-up is a modification of a previously developed apparatus, which has been employed to perform a typical JKR experiment~\cite{Leger1998}. 
A spherical lens is fixed on the lower surface of a glass plate, which is coupled to a motorized linear translation stage (MLTS), allowing the vertical displacement of the lens.
The displacement of the MLTS is measured by a resistive displacement sensor.
The spherical lenses were made by moulding and crosslinking of PDMS (Dow Corning, Sylgard 184, 10/1 w/w) using a well-established technique~\cite{Poulard2011}.
In turn, a patterned substrate is placed over the upper surface of a solid plate, that is connected to a force sensor.
Two different types of patterned substrates were tested: (1) substrates made with the same cross-linked PDMS elastomer as the lens, leading to the intermediate relative condition described in Table~\ref{Tab:Subs} and (2) Araldite 2020 covered with an absorbed PDMS layer (Sylgard 184 base only), leading to the rigid relative condition described in Table~\ref{Tab:Subs}.

The described experimental set-up allows us to quantify the indentation and the compression force, once the contact between the lens and the substrate is established.
All data reported below correspond to quasi-static approach experiments: Small successive displacement steps of $1\, \mu$m are applied, when the lens is pushed toward the substrate, waiting after each step to fully attain the stationary conditions of contact radius and load.
Additionally, a microscope, in communication with a computer, is placed above the lens and substrate, and aligned with the axis of symmetry of the lens.
Since the glass plate and the spherical lens are transparent, the microscope configuration enables us to determine the radius of the contact area.

As described in Table~\ref{Tab:Subs}, the height $h$ and diameter $d$ of the cylindrical pillars that compose the hexagonal pattern at the substrate has been fixed to a single value.
The only variable geometrical characteristic of the patterned substrate was the pitch $e$ between pillars.
This quantity has been varied in the range $e\in\left[8\, ,\, 57\right]\, \mu$m, which according to Eq.~\eqref{eq:phi} leads to experimental values of the surface fraction in the range $\phi\in\left(0.01\, ,\, 0.29\right)$.
Nevertheless, the contact transition has not been observed for $e<12.5\, \mu$m, which corresponds to $\phi>0.21$.

The normalized critical force $F_c\gamma_{int}/(hd)$, contact radius $a_c/e$ and indentation $\zeta_c/h$ as functions of $\phi$, obtained from the retrieved experimental data, are shown in Fig.~\ref{Fig:FaZPhi}.
The three dependent variables show a trend to increase as the surface fraction $\phi$ grows.
Also, the values of the critical force, contact radius and indentation for the rigid substrate experiments are slightly higher than those of the intermediate substrate rigidity.

\subsection{Numerical simulations}

The numerical technique, introduced in a previous study~\cite{Ledesma2016}, consists in an optimization method to determine the size of the contact radius $a$ and the discrete pressure distribution $\sigma$ over a mesh of square elements, each one subjected to a constant stress~\cite{Love1929,Johnson}.
The details of the numerical method, employed to find a numerical solution for this discrete contact problem, had been addressed in the same previous study~\cite{Ledesma2016}, including the estimation of $F$ and $a$, for a given $\zeta$.
The sole addition in the numerical method consisted in the implementation of an algorithm to progressively increase the indentation $\zeta$, until the gap $\Delta(x,y)$ between the lens and the bottom substrate becomes zero at some localized positions $(x_c,y_c)$, indicating that the critical values of the indentation $\zeta_c$ has been attained.

According to the system of coordinates depicted in Fig.~\ref{Fig:Schema}, the lowest point of the lens has been aligned with a target position atop the textured substrate.
Simulations had been performed, for the reduced distances $\delta x/e=\left\{0,1/6,1/3,1/2,2/3,\sqrt{3}/2\right\}$ along the $x$ axis, where $\delta x$ corresponds to a displacement of the target position atop the textured substrate.
In other words, the target position has been changed from the center of the 1st pillar up to the intermediate position between two adjacent pillars.
This procedure has been done to diminish the deterministic consequences of the lens and textured substrate relative position, which are significant for small values of the surface fractions $\phi$.

From the simulation results, the normalized critical force $F_c\gamma_{int}/(hd)$, contact radius $a_c/e$ and indentation $\zeta_c/h$ are retrieved as functions of $\phi$, and presented in Fig.~\ref{Fig:FaZPhi}.
As expected, for small values of $\phi<0.003$, the numerical results depict behaviors that are completely dependent of the relative positions of the lens and substrate, mainly observed as different curves for the force and contact radius.
For $\delta x\rightarrow 0$, the lens makes contact with the bottom substrate at a position $r\ll e$, leading to relatively small values of the contact radius $a_c/e\sim0.01$, whereas the critical force is $F_c\sim hd/\gamma_{int}$.
For increasing values toward $\delta x\rightarrow(\sqrt{3}/2)e$, the lens touches the bottom substrate with progressively less effort, provoking smaller values of the critical force $F_c< hd/\gamma_{int}$.
The mixed contact position, at which contact between the lens and the bottom substrate occurs, is located very close to $r=0$, but the lens makes contact with the top of the off-center 1st pillar at $x\sim\delta x$, thus making $a_c\sim\delta x$.
This deterministic disparity weakens as $\phi$ increases and the different curves become indistinguishable for $\phi\geq0.01$, as both variables $F_c$ and $a_c$ increase as $\phi$ becomes larger.
In turn, the curves for the indentation are almost identical regardless of the deterministic conditions, a monotonic growth of $\zeta/h$ as $\phi$ increases is observed.
For the three aforementioned critical variables, the orders of magnitude and the trends of the numerical results are in very good agreement with the experimental results.

\subsection{Theoretical analysis}


The theoretical approach is based on the Boussinesq-Cerruti solution for uniform displacement over a circular region~\cite{Johnson,Sneddon1945} to describe the displacement fields due to a single pillar-lens contact.
Considering the $n$th pillar, the aforementioned solution implies the combined displacement field of the lens and the textured substrate:
\begin{align}
w_n\left(x,y\right)&=w_{l,n}\left(x,y\right)+w_{s,n}\left(x,y\right) \notag \\
&=\begin{cases} \dfrac{\gamma_{int}F_n}{d} & \text{if } r_n\leq d/2 \ , \\
\dfrac{\gamma_{ext}F_n}{d}\arcsin\left(\dfrac{d}{2 r_n}\right)
& \text{if } r_n> d/2 \ , \end{cases}
\label{eq:BCdispl}
\end{align}
with the introduction of the following parameters:
\begin{subequations}
\begin{align}
\gamma_{int}&=\dfrac{1-\nu_l^2}{E_l}+\dfrac{1-\nu_s^2}{E_s}+\dfrac{4 h}{\pi\, E_s\, d} \ , \\
\gamma_{ext}&=\dfrac{2}{\pi}\left(\dfrac{1-\nu_l^2}{E_l}+\dfrac{1-\nu_s^2}{E_s}\right) \ .
\end{align}
\end{subequations}
Also in Eq.~\eqref{eq:BCdispl}, the change of height of the pillar has been approximated by the Hooke's law, and $F_n$ is the force mutually exerted between the lens and the $n$th pillar.
According to the Boussinesq-Cerruti solution, we also know that:
\begin{align}
F_n&=\dfrac{\pi d^2}{2}\sigma\left(r_{n,0}\right) \ ,
\label{eq:BCforce}
\end{align}
where $\sigma\left(r_{n,0}\right)$ is the stress at the center of the $n$th pillar and $r_{n,0}$, obtained as $r_{n,0}=r_{n}\left(0,0\right)$ in Eq.~\eqref{eq:radd},  its location relative to the origin of coordinates.

For this analysis, we consider that the target position is placed over the center of the 1st pillar (see Fig.~\ref{Fig:Schema} and set $\delta x=0$), which makes $(x_1,y_1)=(0,0)$.
Additionally, from previous experience~\cite{Ledesma2016}, we can imply that the stress distribution should follow a Hertzian-like behavior:
\begin{equation}
\sigma\left(r_{n,0}\right)=\sigma_0\sqrt{1-\left(\dfrac{r_{n,0}}{a}\right)^2} \ ,
\label{eq:Hstress}
\end{equation}
with $\sigma_0$ being the stress at the center of the central pillar.

Considering all the quantities at the origin, $(x,y)=(0,0)$, where the gap is always $\Delta\left(0,0\right)=0$, we start by combining Eqs.~\eqref{eq:ind} and \eqref{eq:totdisp}.
Injecting Eqs.~\eqref{eq:BCdispl}--\eqref{eq:Hstress}, considering that for large relative distances, \emph{i.e.}, $r_{n,0}>d/2$, the inverse trigonometric function in each term can be approximated by $d/(2 r_{n,0})$, introducing the constant $\eta=\left[\sqrt{3}/(2\pi)\right]^{1/2}$ and employing the definition of $\phi$ given in Eq.~\eqref{eq:phi}, leads to an expression that relates the indentation, the center stress and the contact radius:
\begin{equation}
\zeta\approx\dfrac{\pi d\sigma_0}{2}\left[\gamma_{int}+\gamma_{ext}\, \eta\, \sqrt{\phi}\, G_0\left(\dfrac{a}{e}\right)\right] \ ,
\label{eq:indApp}
\end{equation}
where we define the auxiliary function $G_0$, which reads:
\begin{equation}
G_0\left(\dfrac{a}{e}\right)=\sum_{n=2}^N \left[\left(\dfrac{e}{r_{n,0}}\right)\sqrt{1-\left(\dfrac{r_{n,0}}{e}\vphantom{\dfrac{e}{r_{n,0}}}\right)^2\left(\dfrac{e}{a}\vphantom{\dfrac{e}{r_{n,0}}}\right)^2}\right] \ .
\label{eq:auxG0}
\end{equation}

In turn, the total force is given by the superposition of the individual forces mutually exerted between the $n$th pillar and the lens:
\begin{equation}
F=\sum_{n=1}^N F_n=\dfrac{\pi d^2 \sigma_0}{2}\, G\left(\dfrac{a}{e}\right) \ ,
\label{eq:fapp}
\end{equation}
with the auxiliary function $G$ defined as:
\begin{equation}
G\left(\dfrac{a}{e}\right)=\sum_{n=1}^N \left[\sqrt{1-\left(\dfrac{r_{n,0}}{e}\vphantom{\dfrac{e}{r_{n,0}}}\right)^2\left(\dfrac{e}{a}\vphantom{\dfrac{e}{r_{n,0}}}\right)^2}\right] \ .
\label{eq:auxG}
\end{equation}
The auxiliary functions $G_0$ and $G$ have been defined in terms of the distances $r_{n,0}$ and $a$, both normalized by the pitch $e$, which allows us to compute both functions only once.
The detailed development that leads to Eqs.~\eqref{eq:indApp} and \eqref{eq:fapp}, and the behaviors of the corresponding auxiliary functions, are presented in the supplemental material~\cite{SuppMat}.

For indentations above $\zeta_H=d^2/\left(4 R\right)$ and its corresponding force $F_H\sim\sqrt{R \zeta_H^3}$, which have been obtained considering Hertzian contact with a contact radius equal to $d/2$, the contact region spans over the whole lid of the central pillar, with $\Delta\left(r\leq d/2\right)=0$, and beyond.
Indentations and forces larger than $\zeta_H$ and $F_H$ provoke either top or mixed contact situations, depending on the coupled response of the lens and the textured substrate.


Now, we assume that the mixed contact position $(x_c,y_c)$ takes place close to the 1st and central pillar, which is located at $(x_1,y_1)=(0,0)$.
Therefore, the following analysis concerns positions $(x,y)$ where a possible contact between the lens and the bottom substrate may occur, for which $r<e$, with the distance $r=r_1=\sqrt{x^2+y^2}$ as in Eq.~\eqref{eq:radd}.
We also use an approximation to the second order of Eq.~\eqref{eq:lens0} around $r=0$, since $r<e$ and $e\ll R$, and we further restrict $(x_c,y_c)$ to locations near the $x$ axis (or their equivalent positions due to the hexagonal symmetry of the system).
Additionally, we assume that $r$ is small compared to $r_{n,0}$.
Once more, the Hertzian-like behavior for the stress distribution is employed, and let's consider large relative distances $r_n>d/2$ to the $n$th pillar, allowing us to approximate the inverse trigonometric function in Eq.~\eqref{eq:BCdispl} by $d/(2 r_n)$.
Even though these simplifications makes us dismiss information about the precise geometry of the deformed configuration of the lens and the textured substrate, a good estimation of the mixed contact conditions are expected.

Applying the aforementioned simplifications, after some math, the employment of Taylor series up to the second order in $r$, and the use of the cosine law to relate some distances, the gap between the lens and the bottom substrate becomes:
\begin{eqnarray}
\Delta(r) &\approx&h
+\dfrac{r^2}{2R}
+\dfrac{\pi d^2\sigma_0}{4}\left\{\gamma_{ext}\left[\dfrac{1}{r} \right.\right. \nonumber\\
&& \left.\left.+\left(\dfrac{2\eta\, \sqrt{\phi}}{d}\right)^3 G_1\left(\dfrac{a}{e}\right)\dfrac{r^2}{2}\right]
-\gamma_{int}\right\} \ ,
\label{eq:gapext3}
\end{eqnarray}
where the auxiliary function $G_1$ is defined as:
\begin{eqnarray}
G_1\left(\dfrac{a}{e}\right) &=\displaystyle\sum_{n=2}^N& \left\{
\left[3\left(\dfrac{x_n}{e}\vphantom{\dfrac{e}{r_{n,0}}}\right)^2\left(\dfrac{e}{r_{n,0}}\right)^2-1\right]\right. \nonumber \\
&& \left.\times\left(\dfrac{e}{r_{n,0}}\right)^3
\sqrt{1-\left(\dfrac{r_{n,0}}{e}\vphantom{\dfrac{e}{r_{n,0}}}\right)^2\left(\dfrac{e}{a}\vphantom{\dfrac{e}{r_{n,0}}}\right)^2}\right\} \ .
\label{eq:auxG1}
\end{eqnarray}
The detailed development that leads to Eqs.~\eqref{eq:gapext3} and the behavior of the corresponding auxiliary function, are presented in the supplemental material~\cite{SuppMat}.

When the mixed contact is triggered, the gap becomes $\Delta\left(r_c\right)=0$, at the positions of first contact between the lens and the bottom substrate $r_c$.
Under these circumstances, the stress $\sigma_0$, the contact radius $a$ and the indentation $\zeta$, displayed in Eqs.~\eqref{eq:indApp} and \eqref{eq:gapext3}, turn into their critical values $\sigma_{0,c}$, $a_c$ and $\zeta_c$, respectively.
The mixed contact position $(x_c,y_c)$ also corresponds to a minimum, which is described by the zero value of the derivative $\Delta_r(r_c)=0$.
From Eq.~\eqref{eq:gapext3} and its derivative with respect to $r$, the system formed by $\Delta(r_c)=0$ and $\Delta_r(r_c)=0$ must be solved in order to determine the critical conditions, that is: the first mixed contact position $r_c=\sqrt{x_c^2+y_c^2}$, the critical stress $\sigma_{0,c}$ and the critical contact radius $a_c$.
The knowledge of these quantities allows the determination of the corresponding critical indentation $\zeta_c$, through Eq.~\eqref{eq:indApp}.
Unfortunately, this nonlinear system of equations remains underdetermined, since an \emph{equation-of-state} defining the behavior of $a_c$ with the other critical conditions remains unknown.

Nevertheless, playing around with both equations allows us to find a polynomial in terms of the mixed contact distance $r_c$: 
\begin{eqnarray}
\left[\left(\dfrac{2\eta\, \sqrt{\phi}}{d}\right)^3G_1\left(\dfrac{a_c}{e}\right)
+\dfrac{2}{Rhd}\left(\dfrac{\gamma_{int}}{\gamma_{ext}}\right)\right]r_c^3 && \nonumber\\
-\left(\dfrac{3}{2Rh}\right)r_c^2
-1
&=& 0 \ ,
\label{eq:rc}
\end{eqnarray}
from which its value can be deduced, with the previous knowledge of $a_c$.
As well, we find the following expression for the critical force $F_c$:
\begin{equation}
F_c=\dfrac{2r_c^3}{\gamma_{ext} R}\left[\dfrac{G\left(\dfrac{a_c}{e}\right)}{1
-\left(\dfrac{2\eta\, \sqrt{\phi}}{d}\right)^3G_1\left(\dfrac{a_c}{e}\right)r_c^3}\right]\ ,
\label{eq:Fc}
\end{equation}
which may be useful once $\sigma_{0,c}$, $r_c$, and $a_c$ are known.

In the next subsections, we make some hypothesis on the magnitude of $a_c$, in order to delimit some possible solutions for $r_c$ and $F_c$.

\subsubsection{Small contact radius $a_c<e$}
\label{sec:small}

For $a_c<e$, only the central pillar $n=N=1$ is in contact with the lens and, thus, the initial condition is that $a_c=d/2$ before the mixed contact is observed.
Moreover, the auxiliary functions $G_0$, $G$, and $G_1$, set forth in Eqs.~\eqref{eq:indApp}, \eqref{eq:fapp} and \eqref{eq:gapext3} to calculate the indentation $\zeta$, the total force $F$, and the gap $\Delta$, respectively, take the values $G_0=0$, $G=1$, and $G_1=0$.
With this assumptions, the introduction of the characteristic distance $L_c$ and ratio $\mu_c$, defined as:
\begin{align}
L_c&=\left[\dfrac{Rhd}{2}\left(\dfrac{\gamma_{ext}}{\gamma_{int}}\right)\right]^{1/3} \ , & \mu_c&=\dfrac{L_c^2}{2Rh}
\label{eq:char}
\end{align}
and, from a first-order approximation around $r_c\rightarrow L_c$, the critical values of the force $F_c$, contact radius $a_c=r_c$ and indentation  $\zeta_c$ are estimated with the following expressions:
\begin{subequations}
\begin{align}
F_c&\approx\dfrac{hd}{\gamma_{int}}\left(1+3\mu_c+O\left\{\mu_c^2\right\}\right) \ , \\
a_c&\approx L_c\left(1+\mu_c+O\left\{\mu_c^2\right\}\right) \ , \\
\zeta_c&\approx h\left(1+3\mu_c+O\left\{\mu_c^2\right\}\right) \ .
\end{align}
\label{eq:small}
\end{subequations}
This solution describes the very first mixed contact conditions as long as a single pillar interacts with the lens.
This requirement is fulfilled when the pitch $e$ remains large enough, which befalls if the surface fraction is:
\begin{equation}
\Phi\lesssim\dfrac{\pi}{2\sqrt{3}}\left(\dfrac{1}{2}+\dfrac{\sqrt{2Rh}}{d}\right)^{-2} \ .
\end{equation}
For the cases that are studied in this work, the single-pillar solution seems to be valid for $\phi\lesssim0.003$.
Equations~\eqref{eq:small} describe fixed values of $F_c$, $a_c$ and $\zeta_c$ that do not depend on $\phi$, which are depicted in Fig.~\ref{Fig:FaZPhi}.
The normalized critical force $F_c\gamma_{int}/(hd)$ and indentation $\zeta_c/h$ appear as horizontal lines in their corresponding graphs, which values are almost identical to the numerical solution for $\phi<0.003$ and an aligned lens-substrate configuration with $\delta x=0$.
Even though $a_c$ presents a constant value in the small contact radius regime, the normalized corresponding variable $a_c/e$ shows a dependence on $\sqrt{\phi}$ due to the relationship described by Eq.~\eqref{eq:phi}.

\begin{figure}
\centering
\includegraphics[width=0.6\textwidth]{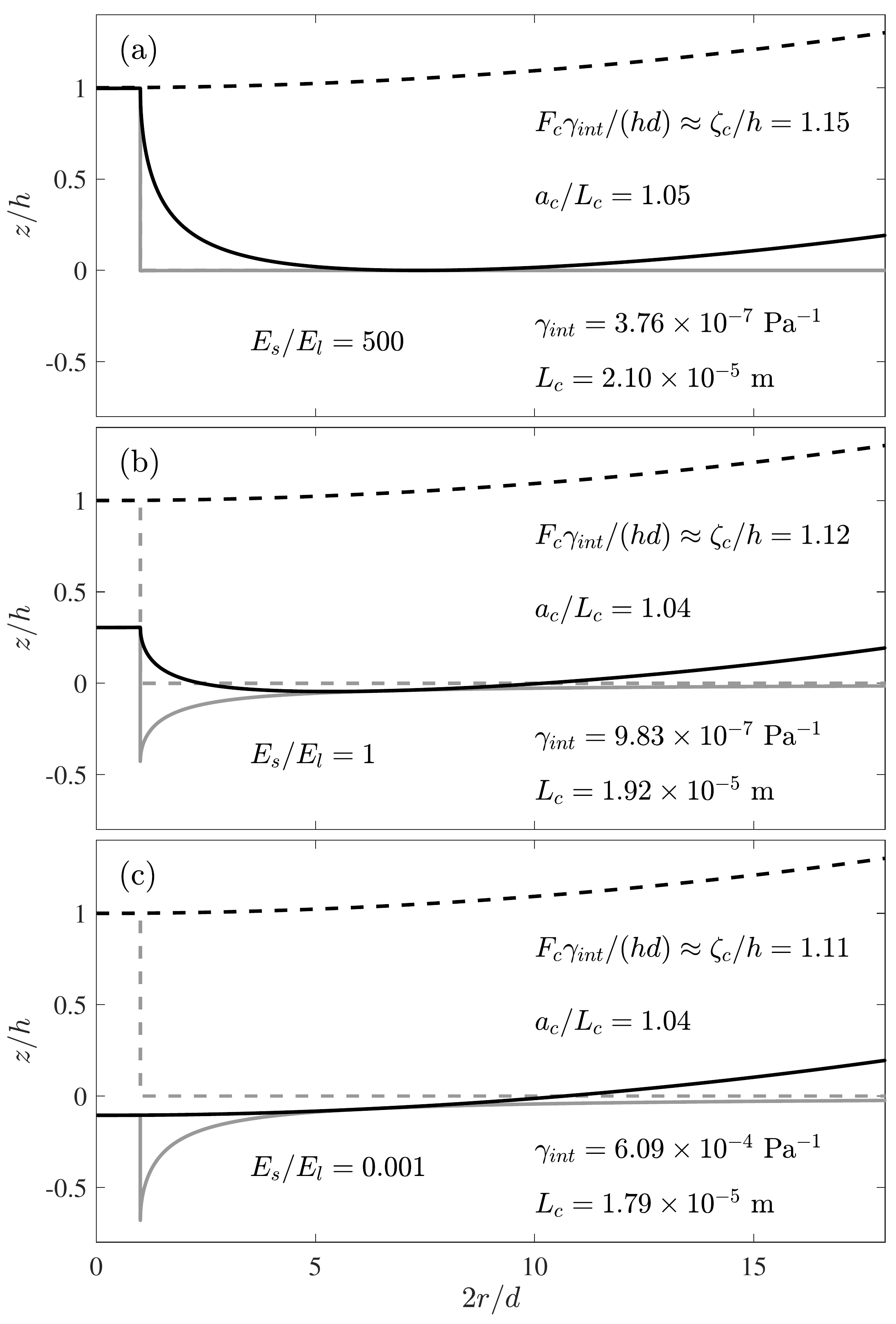}
\caption{
Deformation profiles for a single-pillar substrate (gray) and lens (black).
The Young's modulus of the substrate $E_s$ varies, leading to three different conditions: (a) rigid, (b) intermediate, and (c) soft substrates.
The three cases correspond to the geometric parameters detailed in Tables~\ref{Tab:Lens} and \ref{Tab:Subs}.
Dashed curves ({\protect\tikz[baseline=-0.6ex]{\protect\draw[ultra thick,black!100!white,dashed](0,0) -- (0.6,0);}}) depict the configuration of the system at the initial top contact conditions (single point contact at $r=0$), whereas solid curves ({\protect\tikz[baseline=-0.6ex]{\protect\draw[black!100!white,line width=1.2pt](0,0) -- (0.6,0);}}) indicate the configuration of the system at the critical contact conditions (transition from top to mixed contact).
}
\label{Fig:rZ}
\end{figure}

Additionally, since the small contact radius condition also corresponds to an axisymmetric problem, the theoretical configuration for the single-pillar first contact between the lens and the textured substrate is presented in Fig.~\ref{Fig:rZ}.
The three substrate relative conditions (rigid, intermediate and soft), detailed in Table~\ref{Tab:Subs}, are given in order to compare their behavior.
For the rigid substrate $E_s/E_l=500$, the deformation is concentrated in the lens, and a relatively large critical force $F_c\sim10^{-5}$ N is required to achieve mixed contact.
For the soft substrate $E_s/E_l=0.001$, the deformation is concentrated in the substrate, and a relatively small critical force $F_c\sim10^{-8}$ is observed.
For the intermediate case $E_s/E_l=1$, the deformation is equally shared between the lens and the substrate, while the critical force is in the same order of magnitude but smaller than half the value of the rigid case.
For the three cases, the critical contact radius and indentation are in the same order of magnitude, suffering a slight and almost uniform increase as the rigidity of the substrate increases.

\subsubsection{Large contact radius $a_c>e$}
\label{sec:large}

For $a_c>e$, there should be $N>1$ pillars in contact, whose positions are such that $r_{n,0}\leq a_c$ for $n=1,2,\dots,N$.
Under this situation, the auxiliary functions $G_0$, $G$ and $G_1$ are fairly described by the following linear, quadratic, and rational functions:
\begin{subequations}
\begin{align}
G_0\left(\dfrac{a_c}{e}\right)&\approx C_{01}\left(\dfrac{a_c}{e}-1\right)\ , \\
G\left(\dfrac{a_c}{e}\right)&\approx1+ C_{1}\left[\left(\dfrac{a_c}{e}\right)^2-1\right]\ , \\
G_1\left(\dfrac{a_c}{e}\right)&\approx C_{11}\left(1-\dfrac{e}{a_c}\right)\ ,
\end{align}
\end{subequations}
respectively, with the coefficients $C_{01}=5.770$, $C_1=2.421$, and $C_{11}=5.519$.
Additionally, for $a_c\geq e$, we assume that the contact radius is a function of the surface fraction:
\begin{align}
a_c&=\sqrt{e^2+\phi Rh} \ ,
\end{align}
where we have also considered that $\zeta_c\sim h$.
Additionally, the previous expression indicates that for small values of $\phi$, we may find that $a_c\approx e$, whereas for large values of $\phi$, the contact radius should be $a_c\approx\sqrt{\phi R\zeta_c}$, according to previous observations~\cite{Ledesma2016}. 
Now, considering a first-order approximation around $\phi\rightarrow 0$, the critical values of the force $F_c$, contact radius $a_c$ and indentation  $\zeta_c$ are estimated with the following expressions:
\begin{subequations}
\begin{eqnarray}
F_c&\approx&\dfrac{hd}{\gamma_{int}}\left(1+3\mu_c+O\left\{\mu_c^2\right\}\right) \nonumber \\
&&\qquad\times
\left(1+C_1k^2\phi^2+O\left\{\phi^{7/2}\right\}\right) \ , \\
a_c&\approx& e\sqrt{1+\left(k \phi\right)^2} \ , \\
\zeta_c&\approx& h\left(1+3\mu_c+O\left\{\mu_c^2\right\}\right) \nonumber \\
&&\times
\left(1+\beta k^2\phi^{5/2}+O\left\{\phi^{7/2}\right\}\right) \ ,
\end{eqnarray}
\label{eq:large}
\end{subequations}
where $k=4\eta^2Rh/d^2$ and $\beta=\eta C_{01}\gamma_{ext}/(2\gamma_{int})$.
This solution describes the very first mixed contact conditions for small and intermediate values of the surface fraction $\phi$, where a significant number of pillars interact with the lens.

For the cases that are studied in this work, the solution for large contact radius seems to be a good approximation in the range $0.003<\phi\lesssim0.2$.
Equations~\eqref{eq:large} describe the behaviors of the load $F_c$, the contact radius $a_c$ and the indentation $\zeta_c$ as functions of $\phi$, all of them presenting a monotonic growth as $\phi$ is increased, as it is depicted in Fig.~\ref{Fig:FaZPhi}.
The aforementioned large-area theoretical behaviors show good agreement for $F_c$ and $a_c$, whereas the trend of $\zeta_c$ presents an important deviation from the numerical results (overestimation), but still remaining within the same order of magnitude, for surface fractions around $\phi\sim0.1$.
An increase of the number of terms taken into account in Eq.~\eqref{eq:large} may lead to a better approximation of $\zeta_c$.

\section{Discussion}

\begin{figure}
\centering
\centering\includegraphics[width=0.65\textwidth]{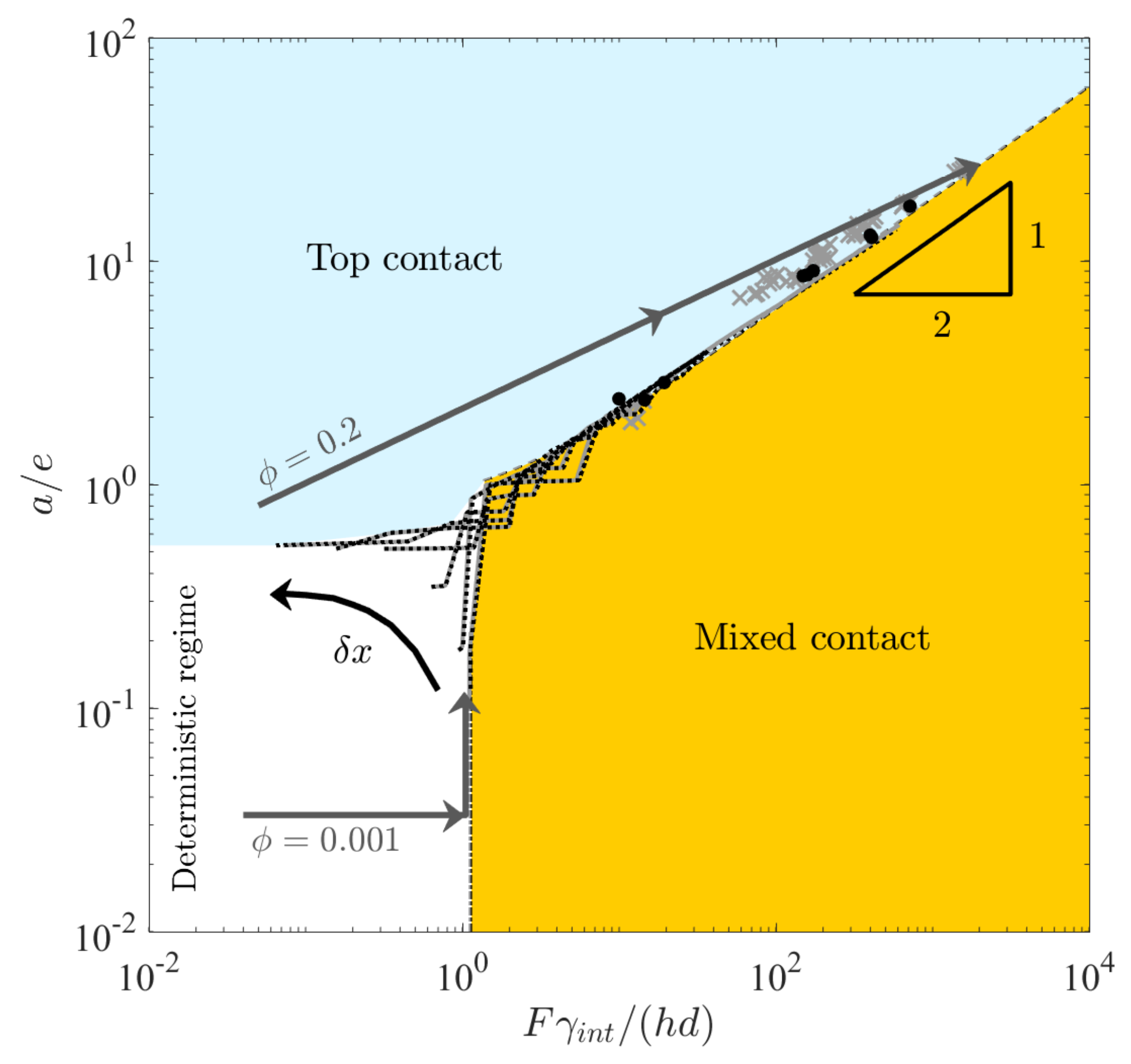}
\caption{
Phase diagram of the contact regime.
The phase space is given in terms of the normalized force $F \gamma_{int}/(hd)$ and the reduced contact radius $a/e$, for the parameters given in Tables~\ref{Tab:Lens} and \ref{Tab:Subs}.
The top and mixed contact regions are bounded by the critical values of the force $F_c$ and the contact radius $a_c$, which were obtained experimentally ($\bm{\times},\bm{\bullet}$) or numerically ({\protect\tikz[baseline=-0.6ex]{\protect\draw[black!100!white,line width=1.2pt](0,0) -- (0.6,0);}},{\protect\tikz[baseline=-0.6ex]{\protect\draw[black!100!white,dotted,line width=1.2pt](0,0) -- (0.6,0);}}) for intermediate (gray,$\bm{\times}$,{\protect\tikz[baseline=-0.6ex]{\protect\draw[black!100!white,line width=1.2pt](0,0) -- (0.6,0);}}) and rigid (black,$\bm{\bullet}$,{\protect\tikz[baseline=-0.6ex]{\protect\draw[black!100!white,dotted,line width=1.2pt](0,0) -- (0.6,0);}}) conditions of the textured substrate.
The approximation described by Eqs.~\eqref{eq:small}, dash-dotted curves ({\protect\tikz[baseline=-0.6ex]{\protect\draw[black!100!white,line width=1.2pt,dash dot](0,0) -- (0.6,0);}}), relates $F_c$ and $a_c$ for geometric situations for which $\phi$ is small, whereas the approximation given by Eq.~\eqref{eq:trans}, dashed curves ({\protect\tikz[baseline=-0.6ex]{\protect\draw[black!100!white,line width=1.2pt,dashed](0,0) -- (0.6,0);}}), corresponds to large values of $\phi$.
The black arrow indicates the direction in which the displacement $\delta x$ of the target atop the substrate increases, from 0 to $(\sqrt{3}/2)e$.
The dark gray arrows indicate examples of evolution trends to attain the contact transition, for small ($\phi=0.001$) and large ($\phi=0.2$) values of the surface fraction occupied by pillars.
}
\label{Fig:PhaseDiag}
\end{figure}

A normalized load-contact radius diagram is shown in Fig.~\ref{Fig:PhaseDiag}, in which the top and mixed contact regimes are located.
The critical values of the load $F_c$ and the contact radius $a_c$, obtained from the experiments, numerical simulations and theoretical analyzes and presented in Fig.~\ref{Fig:FaZPhi}, are shown in Fig.~\ref{Fig:PhaseDiag}.
Curves composed of the succession of critical coordinates $(F_c,a_c)$ create a boundary between the two contact regimes.
This boundary is blurry for the combination of values of the normalized force $F \gamma_{int}/(hd)<1$ and the reduced contact radius $a/e<1$, where a deterministic regime dictates the conditions of the transition from top to mixed contact, behavior that comes from small surface fraction configurations $\phi<0.003$.
The boundary is sharper for larger values of $F \gamma_{int}/(hd)\gg 1$ and, from Eqs.~\eqref{eq:large}, one finds that the contact regions are bounded by the curve given by:
\begin{align}
\dfrac{a_c}{e}\approx \left[\dfrac{1}{C_1\left(1+3\mu_c\right)}\left(\dfrac{F_c \gamma_{int}}{hd}\right)\right]^{1/2} \ .
\label{eq:trans}
\end{align}
The experimental critical coordinates $(F_c,a_c)$ are close to the theoretical findings, which allows us to continue the discussion focusing on these trendlines.

At the deterministic region, since $\phi$ is small, the relative position of the lens with respect to the textured substrate will specify whether top contact or mixed contact occurs.
Let us consider a more precise condition such as $\delta x/e=0$, which corresponds to a textured substrate with its first pillar aligned with the lens center.
For a given surface fraction $\phi$, it is well-known that the contact radius is a function of the force $a=a(F)$.
Now, assume that a zero force condition $F=0$ will start with top contact and a contact radius $a=0$.
While increasing $F$, the overall deformation of the system will increase and the contact radius will become $a=d$, which will remain constant until other pillars are reached by the lens, changing the contact radius to $a/e\approx 1$.
This intermittent augmentation of the contact radius will continue as more pillars come into contact, for instance taking the consecutive values of $a/e=\{1,\sqrt{3},2,\sqrt{7},3,2\sqrt{3},\dots\}$.
For small surface fractions $\phi$, this discontinuous process continues until the critical force $F_c$ is attained and the very first mixed contact condition takes place.
The case of $\phi=0.001$ is presented in Fig.~\ref{Fig:PhaseDiag}, where we can observe that the contact radius $a=d$ is generated for forces $F\gamma_{int}/(hd)<1$.
When the force increases to $F=F_c$, given by Eq.~(\ref{eq:small}a), the very first mixed contact occurs and the contact radius jumps to $a=a_c$, given by Eq.~(\ref{eq:small}b), according to the small contact radius $a_c<e$ approximation.
For large values of $\phi$, a significant amount of pillars join rapidly to the top contact and the dependence of $a$ on $F$ becomes Hertzian-like, thus following the trend $a\sim\sqrt[3]{RF}$ until $F\rightarrow Fc$, given by Eq.~(\ref{eq:large}a) , and the transition toward mixed contact happens, with a contact radius $a=a_c$ given by Eq.~(\ref{eq:large}b) .
The general trend of the case $\phi=0.2$ is presented in Fig.~\ref{Fig:PhaseDiag}, where the contact radius increases roughly as $a\approx\sqrt[3]{(3\pi/8)\gamma_{ext}RF}$ until the force reaches $F=F_c$ and the contact radius becomes $a=a_c$, according to the large contact radius $a_c>e$ approximation.
Now, taking into consideration an off-center situation with $\delta x/e\in(0,\sqrt{3}/2]$ and a surface fraction in the range $\phi\in(0.001,0.2)$, depending on the precise value of $\delta_x$, one of the above-mentioned transitions from top to mixed contact may take place.

\section{Conclusions}

In the present work, the transition of the elastic contact regime between a spherical lens and a textured substrate has been studied in depth, performing experiments and numerical simulations, and developing a theoretical approach.
The substrate consists of a hexagonal lattice of pillars over a flat substrate, with the pillars covering a surface fraction $\phi$ of the substrate.

A top contact regime, defined as the contact only between the lens and the top of the pillars, is observed when the compression force $F$ is relatively small.
A mixed contact regime, defined as the simultaneous occurrence of top contact and the contact between the lens and the substrate at some specific positions, is observed when $F$ is relatively large.
The transition between the aforementioned contact regimes takes place when a precise value of $F$ is attained, which is the critical force $F_c$ for a given value of the surface fraction $\phi$.
This critical condition is accompanied by a specific size of the contact region, described by a critical contact radius $a_c$, and a particular magnitude of the indentation $\zeta_c$.

With the use of reduced variables for the critical force $F_c\gamma_{int}/(hd)$, contact radius $a_c/e$ and indentation $\zeta_c/h$, the system behavior and the dependence on the surface fraction $\phi$ is straightforward.
For $\phi\leq0.003$, the transition from top to mixed contact is strongly dependent on the relative position $\delta x$ of the lens over the textured surface.
When $\delta x=0$, a threshold force $F_c\sim hd/\gamma_{int}$, given by the small contact radius approximation, should be exceeded to observe a mixed contact, whereas for $\delta x=(\sqrt{3}/2)e$ it occurs spontaneously with the application of a negligible force.
When $\delta x=0$, the contact radius corresponds to the contact position between the substrate and the lens, described by the single-pillar case or small contact radius approximation, whereas for $\delta x=(\sqrt{3}/2)e$ the contact radius scales as $a_c\sim\delta x$.
Intermediate values of $\delta x$ lead to behaviors between these two limit cases.
For $0.003<\phi<0.2$, all the results (experimental, numerical and theoretical) collapse into general trends and the transition from top to mixed contact only depends on $\phi$.
The critical force behaves as $F_c\propto \left(1+C_1 k^2 \phi^2\right)$, where $C_1$ is a constant and $k$ is a product of geometric parameters of the pillars.
In turn, the contact radius behaves as $a_c\propto\sqrt{1+k^2\phi^2}$ and the indentation as $\zeta_c\propto\left(1+\beta k^2\phi^{5/2}\right)$.

To summarize these results, a phase diagram is depicted in Fig.~\ref{Fig:PhaseDiag} in which axes correspond to the reduced force $F\gamma_{int}/(hd)$ and the normalized contact radius $a/e$.
The top contact and mixed contact regions are shown, bounded by critical curves that define the regime transition.
Additionally a deterministic regime appears for the quadrant where $F\gamma_{int}/(hd)<1$ and $a/e$, for which the relative position of the lens and the textured substrate defines the transition process from top to mixed contact.
This deterministic regime is mainly discernible for systems presenting small surface fractions $\phi<0.003$.
For larger values in the range $\phi\in[0.003,0.2]$, the curves that define the transition from top to mixed contact collapse into a single trend, which is fairly well described by a large contact radius approximation, roughly given by $a_c/e\sim[F_c\gamma_{int}/(hd)]^{1/2}$.

Despite the insight gained with this analysis, the full understanding of the contact mechanics between rough surfaces requires more research to be done.
For instance, the mixed contact between a lens and a textured surface has been studied experimentally~\cite{Crosby2005,Verneuil2007,Hisler2013,Poulard2013,Poulard2015}, under conditions for which adhesion cannot be disregarded.
Therefore, the next step is to incorporate the role of adhesion to the numerical simulation tools that we have developed and perform a rigorous parametric analysis.
Our efforts are currently being focused on this problem, and the corresponding methodology and analysis will be presented in a future publication.

\bibliography{Unhappy_Fakir_v05.bbl}

\end{document}